# Сетевые интернет-коммуникации как инструмент маркетинга
М.Л. Калужский, В.В. Карпов

*Аннотация*: В статье рассматриваются особенности применения сетевых коммуникаций для продвижения товаров в Интернет. Широкое развитие Интернет-технологий превратило социальные коммуникации в самостоятельный инструмент маркетинга. Авторы классифицируют и анализируют возможности использования сетевых интернет-коммуникаций в маркетинговой среде.

*Ключевые слова*: электронная коммерция, маркетинг, интернет-маркетинг, социальные коммуникации, социальные сети, сетевые коммуникации.

# Network Internet-communications as an instrument of marketing
M.L. Kaluzhsky, V.V. Karpov

*Abstract*: The article is about the features of application of network Internet-communications for advancement of the goods. Wide development of Internet technologies has transformed social communications into the independent tool of marketing. Authors classify and analyze possibilities of use of network Internet-communications in the marketing environment.

Бурное развитие сетевых интернет-коммуникаций и электронной коммерции спровоцировало возникновение новых форм продвижения товаров, потенциал развития которых пока ещё не получил должного развития. Сегодня мы являемся свидетелями коренной трансформации методов и приёмов маркетинга, обусловленных особенностями сетевых интернет-коммуникаций.

Это явление не просто открывает новые возможности для продвижения товаров в сети. Сетевые интернет-коммуникации формируют принципиально новую институциональную среду электронной коммерции. Такая среда практически не пересекается с традиционной институциональной средой, располагаясь вне сферы влияния участников традиционных маркетинговых отношений.

Перспективные формы электронной коммерции всегда ассоциировались с развитием компьютерных технологий. Данное обстоятельство объясняется тем, что компьютерные технологии гораздо ближе в институциональном отношении к традиционной экономике, чем виртуальные технологии интернет-коммуникаций. Однако компьютерные технологии лишь определяют условия сетевых технологий и электронной коммерции, но не более того. Эта составляющая сетевой экономики в равной мере доступна всем участникам рынка и потому может учитываться, что называется, «при прочих равных».

Иначе обстоит дело с чисто новыми формами продвижения товаров в сети Интернет. Тут действуют совершенно иные законы рыночного ведения и иные механизмы маркетинговых коммуникаций. Проблема заключается в том, что маркетинговые особенности сетевых интернет-коммуникаций до сих пор не получили должного отражения не только в методологии маркетинга, но и в его теории маркетинга.

Сегодня институциональная сторона сетевых маркетинговых коммуникаций в Интернете продолжает находиться вне поля зрения традиционных социально-экономических институтов. Развитие идёт параллельно, почти не соприкасаясь и не пересекаясь. Здесь доминируют свои институциональные процессы, отражающие становление новых экономических отношений в новых (виртуальных) условиях экономической деятельности.

Отчасти описанная ситуация напоминает фантастический параллельный мир, где нет одного из измерений (географических расстояний) и действуют иные физические законы. Для такого мира неприменима обыденная система координат. Здесь требуется совершенно иной инструментарий для анализа, основанный на понимании законов и реалий новой дей-ствительности. Положение усугубляется тем, что новый виртуальный мир развивается без оглядки на старый мир материального производства. Будучи фактически независимым от материальных факторов, он стремительно изменяется, легко трансформируясь в соответствии с изменениями запросов участвующих субъектов виртуальных взаимоотношений.

В компьютерной среде технологические новации на ближайшую перспективу легко прогнозируются, так как цикл от разработки и внедрения до массового использования компьютерных технологий достаточно длинен. В виртуальной среде и формы сетевых коммуникаций также во многом виртуальны. Это означает, что их возникновение напрямую не зависит от динамики развития компьютерных технологий.

Сетевые формы коммуникаций в Интернете возникают на основе спроса буквально из ничего. Сначала некто, не обладающий значительными материальными ресурсами, предлагает форму взаимоотношений пользователей Всемирной сети, а затем пользователи Интернета наполняют её содержанием, вознося создателя на Олимп сетевого бизнеса. Поэтому экономический анализ маркетингового потенциала сетевых коммуникаций невозможен вне учёта особенностей институциональной среды, определяющей особенности и методы продвижения товаров в сети Интернет. Теория маркетинга тут пока отстает от практики маркетинга.

Сегодня в сети можно обнаружить самые различные формы институционализации коммуникативной среды, которые, однако, вполне поддаются классификации. Многие из этих форм с момента их появления активно используются для продвижения товаров и услуг в сети Интернет. Причем эффективность такого продвижения позволяет говорить о формировании нового маркетингового инструментария, кардинально отличающегося от традиционных методов продвижения и на равных конкурирующего с традиционной практикой маркетинга.

**1. Социальные сети** (*Social Networking Services*) представляют собой интернет-сервисы, предназначенные для организации и поддержания сетевых коммуникаций. Интернет-маркетинг при продвижении продукции здесь ориентируется на следующие маркетинговые возможности социальных сетей:

1. Содержащуюся в профилях пользователей индивидуальную информацию (возраст, место жительства, интересы, образование, хобби и т.д.).

3. Возможность пользователей посещать страницы других пользователей и формировать общественное мнение о товарах, услугах и компаниях.

2. Возможность пользователей делиться между собой полезной информацией и формировать сетевые объединения (группы) по интересам и т.д.

Первая социальная сеть «Classmates» (англ. – «Одноклассники») возникла в США ещё в 1995 году.[1] Однако их бурное развитие началось лишь в 2003-2004 гг. с появлением социальных сетей «MySpace» и «Facebook». Сегодня социальные сети всё больше превращаются в ведущую сферу сетевых интернет-коммуникаций. По данным компании «ComScore» уже в 2011 г. социальными сетями в мире пользовались 1,2 млрд. человек (82% интернет-пользователей).[2]

В России первые успешные социальные сети «Одноклассники» и «Вконтакте» появились в 2006 году. Темпы роста популярности социальных сетей в России идут в русле общемировых показателей. По данным компании «ComScore» в 2011 году 88% пользователей Рунета имели аккаунты в социальных сетях, что лишь на 10% отстаёт от показателей США, Великобритании, Испании и ЮАР.[3]

Общая численность пользовательских аккаунтов российских социальных сетей по данным компании «J'son & Partners Consulting» в 2011 г. составила 43,6 млн. аккаунтов. Из них на социальную сеть «Вконтакте» приходится 28,8 млн. аккаунтов, на «Мой мир» («Mail.RU») – 18,6 млн. аккаунтов, на «Одноклассники.ру» – 18,0 млн. аккаунтов.[4]

---

[1] Сайт социальной сети «Classmates». – http://www.classmates.com.
[2] Приводится по: Интернет в России: Состояние, тенденции и перспективы развития. 2012. Отчет. – С. 52.
[3] Там же.

Первичной ячейкой любой социальной сети является группа (круг и т.п.) – самоорганизующееся сообщество, объединённое общими интересами, целями или потребностями. Поэтому маркетинг в социальных сетях существенно отличается от маркетинга на специализированных сайтах продавцов. Он малорезультативен, когда покупателей убеждают приобрести товар, но даёт впечатляющие результаты при доверительном общении с клиентами.

Продавец, продвигающий в социальных сетях свои товары, сегодня уже не способен диктовать правила и условия участникам сети. Поэтому социальные сети нельзя рассматривать ни в качестве канала сбыта продукции, ни в качестве канала маркетинговых коммуникаций. Эффективность их использования определяется степенью вовлеченности продавца в процесс неформальной институционализации сетевого сообщества.

Для эффективного применения маркетинг в социальных сетях должен стать результатом осознанного выбора аудитории по отношению к товару, бренду или производителю. Лишь в этом случае социальная сеть превращается одновременно в маркетинговую панель, инструмент продвижения и механизм обратной связи с целевой аудиторий.

Сегодня можно выделить как минимум три основных разновидности социальных сетей:

1. *Общеформатные социальные сети*. Такие сети предназначены для неформального общения пользователей («Facebook», «Вконтакте», «Одноклассники», «Мой мир» и др.). Огромная аудитория делает общеформатные сети особенно привлекательными для использования в маркетинговых целях.

Общеформатные сети предоставляют пользователям большой набор маркетинговых инструментов интернет-маркетинга. Они идут в авангарде технического прогресса в этой сфере, предлагаю пользователям широчайший ассортимент интернет-сервисов и мобильных технологий (GPS, Bluetooth, Wi-Fi, Geo-IP и др.).

Так, например, сервис «Таргет@Mail.ru», запущенный компанией «Mail.Ru Group» в 2011 г., предоставляет пользователям социальных сетей «Одноклассники» и «Мой мир» возможность таргетировать рекламу по географическим и демографическим параметрам, а также по времени суток и дням недели.[5] Социальная сеть «Вконтакте» предлагает рекламодателям таргетинг по географическим и демографическим параметрам, социальному положению и сфере деятельности, а также по интересам пользователей.[6]

2. *Профессиональные социальные сети*. Такие сети предназначены для профессионального общения пользователей («Профессионалы.Ру, «RB.ru» и др.). Они идеально подходят для рекрутинга, создания виртуальных бизнес-проектов, а также для продвижения товаров и услуг в профессиональной среде.

Профессиональные сайты также предоставляют пользователям возможности таргетирования рекламы. Ассортимент и возможности такого рода услуг несколько отстают от аналогичных параметров общеформатных сетей. Отчасти это можно объяснить спецификой аудитории, ориентированной на профессиональное общение. Поэтому основными инструментами интернет-маркетинга в профессиональных сетях являются открытые сообщества, онлайн-конференции, блоги и баннерная реклама.

Например, социальная сеть «Профессионалы.Ру» предлагает своим пользователям воспользоваться сервисом микрорекламы для создания и размещения рекламных объявлений на страницах сети.[7] Таргетинг предусматривает выбор целевой аудитории и страниц сайта для размещения рекламы. Стандартные рекламные объявления, состоящие из 100 символов и небольшого (60х60 пикселей) изображения, отсылают посетителей на сайт рекламодателя.

3. *Социальные сети по интересам*. Преимущество социальных сетей по интересам заключается в их уникальности. Такие сети образуются для общения пользователей, объеди-ненных общими интересами. Они представляют собой концентрированную целевую аудиторию для компаний, товары или услуги которых актуальны для аудитории сети. В этом смысле сети по интересам являют собой логическое завершение процесса эволюции тематических форумов.

---

[4] Там же. – С. 53.
[5] Сайт компании «Mail.Ru Group».– https://target.mail.ru.
[6] Сайт социальной сети «Вконтакте». – http://vk.com/ads.
[7] Сайт социальной сети «Профессионалы.Ру». http://professionali.ru/info/adsfaq.

В качестве примера можно привести российскую социальную сеть собаководов «Догстер».[8] Помимо баннерной рекламы, рекламодатели имеют возможность размещать рекламные статьи, спонсировать мероприятия и участвовать в рассылке новостей. Для посетителей работает биржа щенков, есть доска объявлений, блоги и тематические форумы, регулярно организуются выставки собак. Всё сделано для того, чтобы собаковод, единожды зарегистрировавшись на ресурсе, оставался там навсегда.

В электронной коммерции социальные сети способны решать сразу несколько маркетинговых задач:

1. *Прямое общение с покупателями*. Для большинства интернет-продавцов наличие аккаунтов в социальных сетях стало признаком солидности фирмы, использующей возможности прямых контактов с клиентами. Появилось даже новое понятие «*Оптимизация сайта под социальные сети*» (англ. *Social media optimization*), обозначающее комплекс мероприятий, направленных на привлечение посетителей из социальных сетей.

Прямые контакты с целевой аудиторией в социальных сетях позволяют быстро и без посредников получать маркетинговую информацию непосредственно от самих потребителей, эффективно апробировать новые маркетинговые подходы, а также экономить время и деньги на традиционной рекламе. Используя социальные сети, продавцы могут не только самостоятельно продвигать товары, но и привлекать для этого энтузиастов из числа активных участников сетевых коммуникаций. Это выводит на новый уровень сам смысл коммуникативные отношения в маркетинге.

2. *Формирование фан-клубов торговых марок*. Объединения поклонников собственной торговой марки – идеальная среда для продвижения товаров через Интернет. Поэтому сообщества приверженцев торговой марки при минимальных вложениях обладают большей маркетинговой эффективностью в сравнении с традиционными рекламой и PR.

В целом фан-клубы являются следующим этапом в развитии маркетинга в социальных сетях после прямых контактов. Их появление означает консолидацию пользователей под эгидой торговой марки или производителя, превращая сетевые сообщества в инструмент маркетинговой политики. Формирование таких групп происходит в социальных сетях как естественным, так и искусственным путём.

Искусственный путь подразумевает привлечение т.н. «лидеров общественного мнения» в лице популярных блоггеров, журналистов и специалистов по связям с общественностью. Такая работа требует постоянного создания информационных поводов, вербовки и поощрения сторонников и т.д. Однако после превышения критической массы пользователей процесс становится автономным.

3. *Инструмент прямых продаж*. Использование социальных сетей в качестве инструмента организации прямых продаж свойственно в основном малому бизнесу. Некоторые его представители добиваются весьма существенных результатов за счёт доверительности общения внутри социальной сети. Особенность сетевых сообществ заключается в наличии объединяющих людей интересов, реализация которых подразумевает высокий уровень доверительного общения.

Социальные сети – идеальная среда для организации прямых продаж. Двигателем процесса становятся энтузиасты, приглашающие других членов сообщества объединяться в группы по интересам. Например, в только одной из трёх групп социальной сети «Вконтакте» псевдо-шоурума «Shop Daniel» по состоянию на 24.08.2012 было 160.188 членов.[9]

---

[8] Сайт социальной сети «Догстер». – http://cms.dogster.ru/about.
[9] См.: Социальная сеть «Вконтакте». – http://vk.com/showroom_shopdaniel.

В других случаях даже без явных лидеров пользователи могут объединяться для совершения коллективных покупок (аналог потребительской кооперации). Например, в той же социальной сети «Вконтакте» существует немало таких групп (молодые матери, любители джинсовой одежды и т.д.).

В качестве основных тенденций развития социальных сетей в России и в мире можно выделить следующее:

1. обязательное наличие социальной составляющей в социальных сетях;
2. приоритетное развитие специализированных социальных сетей;
3. расширение функциональных возможностей социальных сетей;
4. расширение коммерческой деятельности в социальных сетях;
5. взаимная интеграция социальных сетей.

Следует отметить, что социальные сети не просто инструмент интернет-маркетинга. В социальных сетях возникают совершенно новые, присущие только им формы электронной коммерции. Они выступают в роли как коммуникативной, так и институциональной сред, создающих условия для формирования новых видов предпринимательской деятельности в сети Интернет.

**2. Блоги** (от англ. «we*b log*» – интернет-журнал) представляют собой веб-сайты, содержащие авторские материла владельцев и комментарии пользователей. Отличительная особенность блогов заключается в их публичности и общедоступности. Посетители блогов могут оставлять комментарии и вступать в полемику с владельцами. Это обстоятельства превращает блоги в особую коммуникативную среду, управляемую владельцами и выполняющую одновременно функции электронной почты, новостного канала, веб-форума и чата.

Первопроходцем блогосферы стала американская компания «Pyra Labs», запустившая в 1999 году первый бесплатный сайт «Blogger.com». Сегодня этот в мире лидирующий блоговый сервис принадлежит компании «Google».[10] Блоггеры на нём могут самостоятельно зарабатывать деньги на размещении рекламы с помощью сервиса «AdSense». В платформу сервиса интегрирована технология «Google Friend Connect», позволяющая объединять подписки читателей с разных сайтов.

Крупнейший блогосервис в России «LiveJournal» («Живой Журнал») по состоянию на 30 ноября 2011 г. насчитывал только в русскоязычном сегменте 1,33 млн. аккаунтов.[11] Ежегодно ЖЖ посещают 530 тысяч авторизованных пользователей.[12] Этот сервис позволяет встраивать в свои информационные материалы кнопки «Like» («Facebook»), «Нравится» («Вконтакте») и «+1» («Google»), что значительно ускоряет процесс распространения информации в сети.

Маркетинговая особенность блогов заключается в том, что их владельцы одновременно являются неформальными лидерами общественного мнения. В этом заключается коренное отличие блоггеров от журналистов. Журналисты ориентируются на редакционную политику своего издания и мнение своего работодателя (редактора, владельца и т.д.). Тогда как блоггеры целиком ориентированы на заинтересованную в их самовыражении аудиторию, что является причиной гораздо большей степени доверия к их материалам.[13]

Наибольшую маркетинговую ценность представляют блоггеры с большой аудиторией, так или иначе совпадающей с целевой аудиторией заказчика. Таких блоггеров можно разделить на две основных категории:[14]

1. *Многотысячники* – блоггеры с многотысячной нецелевой аудиторией. Обычно это авторы материалов на общедоступные темы: отдых, политика, воспитание детей и т.п.

---

[10] Интернет-сервис «Blogger» компании «Google». – http://www.blogger.com.
[11] Сайт сетевого сообщества «Живой Журнал». – http://www.livejournal.com.
[12] Интернет в России: Состояние, тенденции и перспективы развития. 2012. Отчет. – С. 55.
[13] Вебер Л. Эффективный маркетинг в Интернете. Социальные сети, блоги, Twitter и другие инструменты продвижения в Сети. – С. 15-16.
[14] Максимюк К.С. Новый Интернет для бизнеса. – С. 129.

2. *Отраслевики* – блоггеры, специализирующиеся на узкоотраслевой тематике. Обычно это сотрудники профильных компаний по личной инициативе или по распоряжению руководителя ведущие свои интернет-дневники.

Многотысячников нецелесообразно привлекать в качестве платных трансляторов коммерчески важной информации. Их главный недостаток заключается в отсутствие профессиональной компетентности в специфике продвигаемой продукции. Вместе с тем, многие их них рассматривают ведение блогов в качестве важного источника доходов.

Отраслевики являются профессионалами в своей отрасли. Поэтому именно они более способны к эффективному продвижению продукции в Интернете. Основная маркетинговая ценность отраслевых блоггеров заключается во влиянии на целевую аудиторию. Единственная проблема состоит в поиске независимых блоггеров, не связанных обязательствами с конкурирующими компаниями.

В качестве методов стимулирования блоггеров далеко не всегда на первое место выходит денежное вознаграждение. Во многих случаях гораздо более эффективными может быть реализация совместных проектов, передача товаров на тестирование и привлечение блоггеров в качестве независимых экспертов. Блогосфера носит в первую очередь некоммерческий характер, а её участниками движут не мотивы, связанные с извлечением прибыли, а мотивы, связанные с самореализацией.[15]

**3. Веб-форумы** представляют собой интернет-приложения, предназначенные для организации общения посетителей на сайте. Отдельный форум состоит из разделов для обсуждения, в которых пользователи создают темы. В рамках тем посетители имеют возможность высказывать своё мнение.

Отклонение от заданной темы обычно запрещено правилами форума и ведёт к удалению сообщения или аннулированию аккаунта. За соблюдением правил следят модераторы (в рамках раздела) и администраторы (в рамках форума), которые могут редактировать, перемещать или удалять сообщения пользователей. В целом форум является саморегулирующимся сообществом пользователей, объединенных общими интересами.

Наибольшее маркетинговое значение имеют два вида веб-форумов:[16]

1. *Корпоративные форумы*, предназначенные для оперативной обратной связи с потребителями и контрагентами. Такие форумы могут быть эффективными только в случае действенного участия в их работе независимых специалистов и энтузиастов. Проблема заключается в низком уровне доверия независимых интернет-пользователей к информации корпоративных ресурсов.

2. *Тематические форумы*, создаваемые энтузиастами с целью объединения пользователей на основе общности интересов. Такие форумы требуют высокого профессионального уровня модераторов. Однако их эффективность гораздо выше из-за большего числа посетителей и повышенного уровня доверия интернет-пользователей.

Ближе всего в традиционной теории маркетинга к понятию «форум» располагаются понятия «панель» и «потребительская конференция», используемые при проведении маркетинговых исследований. Особенность форума заключается в том, что здесь невозможен монолог или даже диалог продавца о достоинствах предлагаемых товаров и услуг. Модератор только задаёт тему для обсуждения, а собеседник в любой момент может присоединиться или выйти из обсуждения.

В целом веб-форумы в силу своей коллективности являются скорее переходной формой к социальным сетям и блогосфере. Форумы не подходят для размещения большого количества технической информации, торговых инструментов и не свойственных сервисов. Поэтому форумы используются обычно для расширенного общения по принципу «вопрос-ответ» с той лишь разницей, что задают вопросы и получают ответы одновременно все пользователи.

---

[15] Там же. – С. 130-135.
[16] Дейнекин Т.В. Интернет-форумы как инструмент маркетинга. – С. 74, 77.

**4. Электронные доски объявлений** представляют собой самую простую и самую доступную форму электронной коммерции с огромным числом участников и минимальным перечнем предоставляемых им услуг. Сайт с объявлениями не содержит торговых инструментов, не несёт ответственности за результаты сделки и часто даже не взимает плату за размещение информационных материалов.

Электронные доски объявлений успешно конкурируют с торговыми площадками по числу зарегистрированных пользователей и числу посещений. Это вынуждает некоторые торговые площадки размещать на своих ресурсах собственные доски объявлений, как например электронная доска объявлений «Slando» на торговой площадке «Молоток.Ру».[17]

Благодаря своей простоте и доступности электронные доски объявлений успешно действуют как в сфере «B2B», так и в сфере «B2C».

В сфере «B2B» наиболее показательным примером может служить Международный онлайн-каталог товаров и услуг «All.Biz», содержащий информацию о коммерческих предложениях 985 тыс. компаний из 67 стран мира.[18] Помимо информации о товарах этот сайт объявлений размещает информацию о торгово-промышленных выставках, выставочных центрах, курсах валют с автоматическим переводом на 26 языков мира включая русский.

В сфере «B2C» наиболее показательным примером может служить запущенная в 2007 году крупнейшая в России электронная доска объявлений «AVITO.ru», содержащая более 6,5 млн. частных объявлений.[19] Помимо тематического поиска сайт предоставляет возможность регионального поиска товаров и услуг. Эта услуга позволяет «AVITO.ru» успешно конкурировать с региональными досками объявлений. Несмотря на то, что сайт ориентирован на частные объявления, он представляет платные услуги по открытию интернет-магазинов на своей платформе.

Кроме того, следует отметить гибридные формы досок объявлений и специализированных информационных сервисов. Такой формой является справочный сервис «2GIS» компании-разработчика электронных справочников ООО «ДубльГИС».[20] По данным компании 2,6 млн. пользователей из более 182 городов России, Украины, Казахстана и Италии регулярно пользуются услугами сервиса. Свежие выпуски 2ГИС выходят ежемесячно в трех версиях: для мобильных телефонов, для персональных компьютеров и онлайн-версия в сети Интернет.

Бесплатный конечный продукт сервиса «2GIS» включает в себя большой массив информации: номера телефонов, адреса, сайты и электронную почту предприятий и организаций, расписание их работы, карт населённых пунктов и маршрутов городского транспорта. Всё это делает «2GIS» гораздо более востребованным, нежели традиционные рекламно-информационные издания.

Разнообразие и масштабность перечисленных выше форм и видов сетевых интернет-коммуникаций лишний раз свидетельствует об огромных возможностях и перспективах развития этой сферы. В институциональном смысле здесь мы имеем дело с типичным результатом сокращения трансакционных издержек. Причем не столько финансовых и материальных издержек, сколько затрат времени и усилий субъектов коммуникативных отношений, поскольку благодаря Интернету скорость и объем межличностных коммуникаций возрастают неимоверно.

Не удивительно, что новые возможности всемирной сети, так или иначе, используются пользователями, в том числе и для маркетингового продвижения товаров. Принципиальное отличие здесь заключается в том, что виртуальная среда Всемирной сети нивелирует естественные коммуникативные ограничения, существующие в повседневной жизни вне Интернета. В Интернете нет расстояний, нет проблем с передачей большого объема не только текстовой, но и аудио-визуальной информации и т.д.

---

[17] Электронная доска частных бесплатных объявлений «Slando». – http://slando.ru.
[18] Международный онлайн-каталог товаров и услуг «All.Biz» – http://www.all.biz.
[19] Сайт бесплатных объявлений «AVITO.ru». – http://www.avito.ru.
[20] Сайт справочного сервиса ООО «ДубльГИС». – www.2gis.ru.

Именно поэтому, казалось бы, даже на первый взгляд не имеющие прямого отношения к маркетингу новые формы интернет-коммуникаций способны превратиться в достаточно эффективный инструмент маркетинга. Причина кроется в мультипликативном эффекте Интернета, позволяющего во много раз увеличить коммуникативные возможности каждого участника сетевых коммуникаций.

Незначительные в обычных условиях маркетинговые возможности межличностного непрофессионального общения при помощи сетевой мультипликации превращаются в мощнейший инструмент продвижения. Это обстоятельство превращает даже самые обычные сетевые коммуникации в один из самых эффективных инструментов маркетинга. Причем в такой инструмент, роль которого постоянно возрастает, а сам он стремительно изменяется, всё время расширяя и углубляя сферу своего применения.